\documentclass[aip,apl,graphicx,amsmath,amssymb,reprint]{revtex4-1}
\usepackage{graphicx}
\usepackage{dcolumn}
\usepackage{bm}

\newcommand{\p}{$\%$}
\newcommand{\pat}{$_{at.}\%~ $}
\newcommand{\che}[1]{$\mathrm{#1}$}

\begin{document}


\title{Fe and N self-diffusion in non-magnetic Fe:N}

\author{M. Gupta}
\email {mgupta@csr.ernet.in}
\author {A. Gupta}
\affiliation{UGC-DAE Consortium for Scientific Research,
University Campus, Khandwa Road, Indore-452 001,India}
\author{R. Gupta}
\affiliation{Acropolis Institute of Technology and Research,
Manglia Square, Indore-453 771, India}
\author{J. Stahn}
\author{M. Horisberger}
\affiliation{Laboratory for Neutron Scattering, ETH Zurich and
Paul Scherrer Institut, CH-5232 Villigen PSI, Switzerland}
\author{A. Wildes}
\affiliation{Institut Laue-Langevin, rue des Martyrs, 38042
Grenoble Cedex, France}

\date{\today}

\begin{abstract}

Fe and N self-diffusion in non-magnetic \che{Fe{:}N} has been
studied using neutron reflectivity. The isotope labelled
multilayers, \che{[Fe{:}N/^{57}Fe{:}N]_{10}} and \che{[Fe{:}N /
Fe^{15}{:}N]_{10}} were prepared using magnetron sputtering. It
was remarkable to observe that N diffusion was slower compared to
Fe while the atomic size of Fe is larger compared to N. An attempt
has been made to understand the diffusion of Fe and N in
non-magnetic \che{Fe{:}N}.
\end{abstract}

\maketitle

Iron nitrides (\che{Fe{:}N}) show a variety of structures and
magnetic properties with a variation in the nitrogen content. With
an increasing atomic percentage (\pat N), their major phases are:
\che{Fe_{16}N_{2}}, \che{Fe_4N}, \che{Fe_3N}, \che{Fe_2N},
\che{FeN} and \che{Fe_3N_4}. For $\leq$25 \pat N, the \che{Fe{:}N}
phases are magnetic.\cite{Wang_JMMM04} A lot of attention has been
driven to \che{\alpha^{\prime\prime}-Fe_{16}N_2} ($\sim$11\,\pat
N) due to the presence of the so-called giant magnetic moment in
this compound.~\cite{Kim:APL:1972,komuro:JAP:1990} Around 20\,\pat
N, \che{\gamma^\prime-Fe_4N} phase is formed which has a
well-defined magnetic properties and crystal
structure.~\cite{PhysRevB.75.2007} Very recently the
\che{\gamma^\prime} phase has received a lot of interest due to
its chemical inertness and mechanically hard surfaces making it a
suitable alternative to pure Fe in magnetic
devices.~\cite{APL.2009,PhysRevB.78.2008,PhysRevB.75.2007,Wit:PRL94,PRL.95.2005}
Between 25-33\,\pat N the \che{Fe{:}N} are known as
\che{\epsilon-Fe_\textit{x}N} ($2\leq x\leq3$), and as N\,\pat
increases from 25\p\, to 33\p, the phase changes from
ferromagnetic \che{Fe_3N} to paramagnetic \che{Fe_2N} at room
temperature. Earlier it was not possible to produce the
\che{Fe{:}N} phases containing more than 33\,\pat N but by using
reactive
sputtering\cite{Schaaf_HypInt95,Gupta:PRB05,PhysRevB.59.1999} and
pulsed laser deposition techniques,\cite{Gupta_JAC01} the
\che{Fe{:}N} phases with $>$33 \pat N were produced. Around
50\,\pat N, the \che{FeN} phases have cubic ZnS and/or NaCl-type
structures which are known as \che{\gamma^{\prime\prime}} and/or
\che{\gamma^{\prime\prime\prime}}. The \che{Fe_3N_4} phase with
even more than 50\,\pat N was predicted by Ching $et~
al.$~\cite{Ching.APL.02}, but has not been evidenced
experimentally.

Recently, the \che{\epsilon-Fe_2N} and
\che{\gamma^{\prime\prime}}/\che{\gamma^{\prime\prime\prime}}
phases have been used as a precursor to prepare
\che{\gamma^\prime-Fe_4N} phase for its use in the spintronic
devices.~\cite{APL.2009,PhysRevB.78.2008,Wit:PRL94} In the present
work, we have prepared single phase \che{\epsilon-Fe_2N} and
\che{\gamma^{\prime\prime\prime}-FeN} compounds and studied the
self-diffusion of Fe and N. A proper understanding of the
stability and nitride formation requires the knowledge of both Fe
and N self-diffusion at atomic length scales. However, there are
no studies on measurements of both Fe and N self-diffusion in
non-magnetic \che{Fe{:}N}. Conventional techniques to measure
self-diffusion (e.g. secondary ion mass spectroscopy, radioactive
tracer etc.) have depth resolutions of several nm. Therefore, to
measure diffusion at nanometer length scale, a technique with
depth resolution in the sub-nm regime is
necessary.~\cite{gupta:PRB04} Here, the method of choice is
neutron reflectometry (NR), which in addition is sensitive to
isotopic contrast: neutron scattering length for natural Fe and
\che{^{57}Fe} are 9.45 and 2.3\,fm, and for natural N and
\che{^{15}N} 9.36 and 6.6\,fm, respectively.


The samples were prepared at room temperature by DC magnetron
sputtering using either a mixture of Ar and N$_{2}$ (samples $A$),
or with pure N$_{2}$ (samples $B$) as sputtering gas. During the
deposition the total gas flow was kept constant at 10\,standard
cm$^{3}$/min (sccm) and the sputtering power at 50\,W. The actual
thicknesses (obtained using NR) and gas flow parameters are;
Samples ($A$): \che{[Fe{:}N(7.5\,nm)/^{57}Fe{:}N(4.5\,nm)]_{10}},
\che{[Fe{:}N(6.4\,nm)/Fe{:}^{15}N(3.2\,nm)]_{10}} multilayers
deposited using sputter gas (\che{N_2 + Ar}) each at 5\,sccm. And
samples ($B$): \che{[Fe{:}N(10.2\,nm)/^{57}Fe{:}N(5.2\,nm)]_{10}},
       \che{[Fe{:}N(10.6\,nm)/Fe{:}^{15}N(5.2\,nm)]_{10}} multilayers using \che{10\,sccm\,
       N_2} as sputter
gas. The multilayer with \che{^{57}Fe} was deposited by
alternatively sputtering Fe/\che{^{57}Fe} enriched targets and the
\che{^{15}N} multilayer was prepared using the same Fe target but
switching between N/\che{^{15}N} gases. A residual gas analyzer
was installed in the sputtering chamber to monitor the isotope
abundance of nitrogen. The samples were characterized using x-ray
diffraction (XRD) using Cu-K$\alpha$ x-rays while thermal
stability was examined using differential scanning calorimetry
(DSC). The local environment of \che{^{57}Fe} atoms was probed
using conversion electron M\"{o}ssbauer spectroscopy (CEMS). The
NR measurements were performed at the reflectometers AMOR at
SINQ/PSI~\cite{Gupta_PramJP04}, and D17 at ILL~\cite{D17.ILL.02},
both in the time-of-flight mode.


\begin{figure} \center
\vspace {-5mm}
\includegraphics [width=70mm,height=60mm]{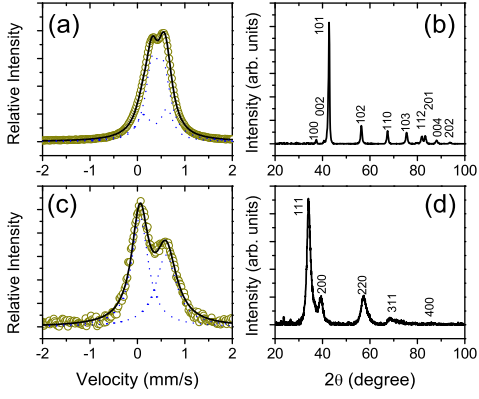}
\vspace {-0.1cm} \caption{\label{fig:fig1} (Color online) CEMS
pattern of sample $A$ (a) and sample $B$ (c). XRD pattern of
sample $A$ (b) and sample $B$ (d).\vspace{-2mm}}
\end{figure}

\begin{figure}  \center
\vspace {-3.5cm}
\includegraphics {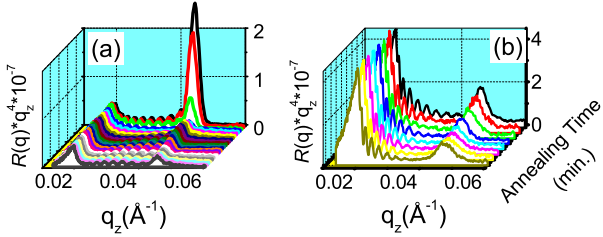}
\vspace {-3cm} \caption{\label{fig:fig2} (Color online) Neutron
reflectivity pattern (reflectivity $R$, multiplied by \che{q_z^4}
for better visuality of Bragg peak) of sample $B$ measured at
463\,K for different annealing time taken with an interval of
30\,min. The time dependence of Fe (a) and N diffusivity (b) can
be seen.\vspace{-5mm}}
\end{figure}

The CEMS and XRD patterns of samples $A$ and $B$ are shown in
fig.~\ref{fig:fig1} (a)-(d). The XRD pattern of sample $A$
(fig.~\ref{fig:fig1}b), shows peaks corresponding to
\che{\epsilon-Fe_\textit{x} N} with hexagonal closed pack
structure. The CEMS spectrum of this sample (fig.~\ref{fig:fig1}a)
shows an asymmetric doublet which was fitted using two doublets
corresponding to Fe-III and Fe-II sites with isomer shifts of
0.43$\pm0.002$ and 0.37$\pm0.002$\,mm/s; quadrupole splitting of
0.24($\pm0.002$) and 0.50$\pm0.004$\,mm/s and the relative area
ratio of 69:31, respectively. The fitted parameters matches well
with the reported values and using the relative area ratio, we
obtain the value of $x=2.23$ following the procedure given in
ref.~\cite{Schaaf_HypInt95}

The XRD pattern of the sample $B$ (fig.~\ref{fig:fig1}d) shows all
the peaks corresponding to \che{\gamma^{\prime\prime\prime}-FeN}
phase with lattice parameter $a=0.454\pm0.001$\,nm. The CEMS
pattern of this sample (fig.~\ref{fig:fig1}c) shows an asymmetric
doublet and was fitted using two singlets with isomer shifts of
$0.05\pm0.002$ and $0.6\pm0.003$ mm/s. The XRD and CEMS parameters
match well with the values obtained by Jouanny
$et\,al.$~\cite{Jouanny2010TSF} for
\che{\gamma^{\prime\prime\prime}-FeN} having ZnS-type structure
with $\approx$50\pat N. The average grain size calculated using
the Scherrer formula for the most intense peak in sample $A$ and
$B$ are 16\,nm and 5\,nm, respectively.

\begin{figure}  \center
\vspace {-5mm}
\includegraphics [width=85mm,height=85mm] {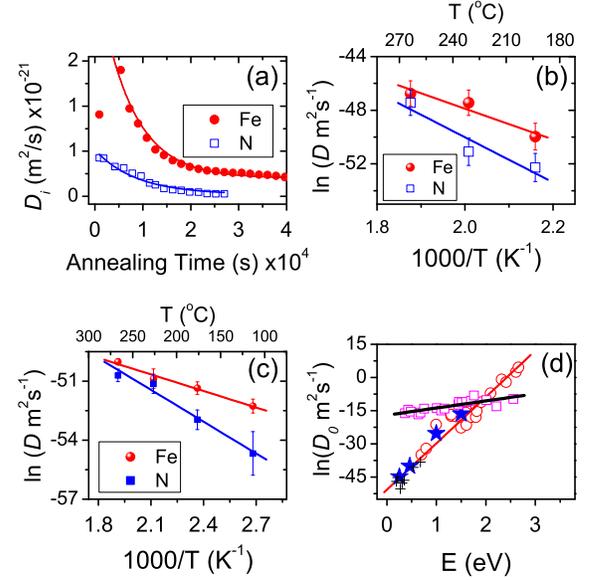}
\vspace {-5mm} \caption{\label{fig:fig3} (Color online) Time
dependence of Fe and N instantaneous diffusivity for sample $B$ at
463\,K (a). Arrhenius behavior of Fe and N diffusivity for sample
$B$ (b) and for sample $A$ (c). The correlation between ln$D_0$
and E (d) for crystalline ($\square$) and amorphous ($\bigcirc$)
alloys~\cite{Faupel_RMP03}. The star ($\bigstar$) represents data
obtained in the present work and ($+$) corresponds to the values
obtain in crystalline thin film
multilayers.~\cite{Wang.PhysRevB.59.10811} \vspace{-5mm}}
\end{figure}

The thermal stability of the samples was studied using XRD and
DSC. In the temperature range of 373-573\,K there was no
appreciable change in the XRD patterns and DSC measurements showed
a strong exothermic peak at 650\,K, indicating out-diffusion of
nitrogen as observed in an earlier study.\cite{gupta:PRB02}
Therefore 523\,K was chosen as the maximum annealing temperature
for the diffusion measurements. The \che{^{57}Fe} and \che{^{15}N}
periodicity gives rise to Bragg peaks in the NR pattern, which
become less sharp as the multilayers are annealed. Some
representative NR patterns for sample $B$ at 463\,K at different
annealing time (with a step of 30\,min) are shown in
fig.~\ref{fig:fig2} (a) and (b) for Fe and N contrast. The pattern
was multiplied with \che{q_z^4} to remove the decay due to
Fresnel's reflectivity. The decay of the Bragg peak intensity can
be used to calculate instantaneous diffusivity ($D_{i}$) using the
expression:\cite{Spapen:APL80} $\ln [I(t)/I(0)] =
-8\pi^{2}n^{2}D_{i} t /d^{2}$ where $I(0)$ is the intensity of the
$n^{th}$ order Bragg peak at time $t = 0$ (before annealing) and
$d$ is the bilayer thickness. As the structure tends to relax, the
time averaged diffusivity is defined by:\cite{Faupel_RMP03}
$\overline{D}=\frac {1} {t} \int_{0}^{t} D_{i} (t^{\prime})~
dt^{\prime}$. Assuming an exponential law for relaxation, $D_i
(t)= \mathrm {Const.}\, \exp (-t/ \tau)+D$, where $D$ is
diffusivity in the relaxed state and $\tau$ is the relaxation
time. Using this relation the diffusivities in the structurally
relaxed state were obtained and are shown in
fig.~\ref{fig:fig3}\,(a). The temperature dependence of
diffusivity obtained in the structurally relaxed state follows an
Arrhenius-type behavior as shown in fig.~\ref{fig:fig3} (b) for
sample $B$. Similarly, for sample $A$, the Arrhenius behavior for
Fe and N diffusivity is shown in fig.~\ref{fig:fig3} (c). The
straight line fit to the data obtained using:
$D=D_{0}\,\exp(-E/k_\mathrm{B}T)$, yields the pre-exponential
factor ($D_{0}$) and activation energy ($E$) which are given in
table~\ref{tab:table1} with $k_\mathrm{B}$ being Boltzman's
constant.

\begin{table}
\caption{\label{tab:table1} Parameters for \che{Fe{:}N} samples.}
\begin{ruledtabular}
\begin{tabular}{lll}
Sample&$A$&$B$  \\  \hline
Composition & \che{\epsilon-Fe_\textit{2.23} N} & \che{\gamma^{\prime\prime\prime}-FeN} \\
\che{E_{Fe}} (eV)& 0.25$\pm0.03$ & 1.0$\pm0.2$ \\
\che{E_{N}}(eV)& 0.46$\pm0.08$ & 1.5$\pm0.3$ \\
ln$D_{0_{Fe}}$(m$^{2}$/s)&-44$\pm1$&-25$\pm10$ \\
ln$D_{0_{N}}$ (m$^{2}$/s)&-41$\pm2$&-16$\pm10$ \\
\end{tabular}
\end{ruledtabular}

 \vspace{-5mm}
\end{table}

It is known that a correlation exists between ln$D_0$ and $E$,
which is followed in all class of materials and indicates the
involved diffusion mechanism.\cite{Faupel_RMP03} While including
our values of $E$ and ln$D_0$ in this correlation
(fig.~\ref{fig:fig3}d), we find that our values follow this
correlation well for diffusion in amorphous alloys. With a
straight line fit of the values obtained for amorphous alloys with
our data, we can calculate the entropy ($\Delta S$) for diffusion
following the procedure given in ref.~\cite{gupta:PRB04} The
calculated values of $\Delta S$ for Fe and N diffusion are
5\,$k_\mathrm{B}$ and 19\,$k_\mathrm{B}$ for sample $A$ and
9\,$k_\mathrm{B}$ and 28\,$k_\mathrm{B}$ for sample $B$,
respectively. In crystalline systems the value of $\Delta S$ is
typically (3-5)\,$k_\mathrm{B}$ corresponding to interstitialcy or
monovacancy mechanism.~\cite{Wang.PhysRevB.59.10811} The
relatively large values of $\Delta S$ indicate that the diffusion
mechanism in the present case is more similar to amorphous alloys
where collective type diffusion occurs involving a group of atoms.
As we see from the XRD measurements, the grain sizes are rather
small (16\,nm and 5\,nm in sample $A$ and $B$). If the diffusion
takes place from the grain boundaries which are not so
well-ordered, the diffusion mechanism will lead to a situation
similar to amorphous alloys.~\cite{Gupta.JNCS.2004} In an earlier
study, we measured the Fe self-diffusion in a nearly equiatomic
amorphous and nonmagnetic FeN alloy~\cite{gupta:PRB02} and found
\che{E_{Fe}}=1.3$\pm0.2$\,eV which is close to
\che{E_{Fe}}=1$\pm0.2$\,eV for sample $B$. Therefore nearly
similar values of $E$ in amorphous and crystalline FeN thin films
indicate a similar type of diffusion mechanism which is
independent of the structure.

The comparison of diffusivity in samples $A$ and $B$ reveals that
the values of $E$ are greater in sample $B$. This indicates a
slower diffusion of Fe or N in sample $B$ as compared to sample
$A$. Looking at the enthalpies of formation (${\Delta H^o}$), the
values for \che{FeN} are the lower as compared to \che{Fe_2N} or
other \che{Fe{:}N} phases.\cite{Tessier_SSS00} This means that
\che{FeN} is expected to be more stable as compared to \che{Fe_2N}
and therefore $E$ for Fe and N self-diffusion should be high in
\che{FeN} as compared to \che{Fe_2N}. Though the differences in
the obtained values of diffusivity in both samples can be
understood in terms of energetics of iron nitrides, it is the
difference between the Fe and N self-diffusion which is counter
intuitive as the atomic size of Fe ($r_\mathrm{Fe} =
0.1274\,\mathrm{nm}$) is larger than N
($r_\mathrm{Fe}/r_\mathrm{N} \approx 1.6$). Therefore, it is
expected that N should be diffusing faster than Fe as was observed
in a N-poor magnetic \che{Fe{:}N}.~\cite{Gupta:AM:2009} In the
literature we wee that atomic size dependency (i.e. smaller atoms
diffuses faster) is observed metal-metal and metal-metalloid
amorphous alloys~\cite{Cahn:JMR:1980,Faupel_RMP03}. However, in
case of phosphorus self-diffusion in
\che{Fe_{40}Ni_{40}P_{14}B_{6}} and in
\che{Pd_{43}Cu_{27}Ni_{10}P_{20}} metallic glass the diffusion
coefficient of P was found to be smaller compared to Fe and
Cu.~\cite{Yamazaki:DSL05:Porc,Valenta:JPSB:1981} It was suggested
that the local chemical interactions around P (strong covalent
bonds) are more important for the diffusion than the atomic size
dependence of constituent elements in the alloy. In the periodic
table N and P are in the same group (VB) and the anomaly observed
for P self-diffusion can be extended to N self-diffusion in
nonmagnetic Fe:N phases. However, as the temperature increases we
observed that the difference between Fe and N diffusivity becomes
smaller.

In conclusion, we measured Fe and N self-diffusion in non-magnetic
\che{\epsilon-Fe_{2.23}N} and
\che{\gamma^{\prime\prime\prime}-FeN} compounds. The activation
energy of Fe and N is higher in
\che{\gamma^{\prime\prime\prime}-FeN} than in
\che{\epsilon-Fe_{2.23}N}. However, the diffusivity of N is
smaller compared to Fe in both samples. The differences in Fe and
N diffusivity can not be understood in terms of atomic seize of Fe
and N.

We acknowledge DST for providing financial support to carry out NR
experiments under its scheme `Utilization of International
Synchrotron Radiation and Neutron Scattering facilities'. A part
of this work was performed under the Indo Swiss Joint Research
Programme with grant no. INT/SWISS/JUAF(9)/2009.


%

\end{document}